\documentclass[aps,prl,twocolumn,superscriptaddress,citeautoscript]{revtex4}
\usepackage{amsfonts}
\usepackage{amssymb}
\usepackage{amsmath}
\usepackage{epsfig}
\usepackage{graphicx}
\usepackage{dcolumn}
\usepackage{bm}
\usepackage{flafter}
\usepackage{float}
\usepackage{color}
\newcommand{\CAS}{CeAuSb$_2$}

\begin{document}

\title{Effect of uniaxial stress on the magnetic phases of CeAuSb$_2$}
\author{Joonbum Park}
\email{jbpark0521@gmail.com}
\affiliation{Max Planck Institute for Chemical Physics of Solids, N\"{o}thnitzer Stra{\ss}e 40, 01187 Dresden, Germany}
\affiliation{Max Planck POSTECH Center for Complex Phase Materials, Pohang University of Science and Technology, Pohang 37673, Republic of Korea}
\author{Hideaki Sakai}
\affiliation{Department of Physics, Osaka University, Toyonaka, Osaka 560-0043, Japan}
\affiliation{PRESTO, Japan Science and Technology Agency, Kawaguchi, Saitama 332-0012, Japan}
\author{Andrew P. Mackenzie}
\affiliation{Max Planck Institute for Chemical Physics of Solids, N\"{o}thnitzer Stra{\ss}e 40, 01187 Dresden, Germany}
\affiliation{Scottish Universities Physics Alliance (SUPA), School of Physics and Astronomy, University of St. Andrews,
St. Andrews KY16 9SS, United Kingdom}
\author{Clifford W. Hicks}
\email{hicks@cpfs.mpg.de}
\affiliation{Max Planck Institute for Chemical Physics of Solids, N\"{o}thnitzer Stra{\ss}e 40, 01187 Dresden, Germany}
\date{\today}

\begin{abstract}
We present results of measurements of resistivity of \CAS{} under the combination of $c$-axis magnetic field
and in-plane uniaxial stress. In unstressed \CAS{} there are two magnetic phases. The low-field A phase is a
single-component spin-density wave (SDW), with $\mathbf{q} = (\eta, \pm \eta, 1/2)$, and the high-field B phase
consists of microscopically coexisting $(\eta, \eta, 1/2)$ and $(\eta, -\eta, 1/2)$ spin-density waves.
Pressure along a $\langle 100 \rangle$ lattice direction is a transverse field to both of these phases, and so
initially has little effect, however eventually induces new low- and high-field phases in which the principal
axes of the SDW components appear to have rotated to the $\langle 100 \rangle$ directions. Under this strong
$\langle 100 \rangle$ compression, the field evolution of the resistivity is much smoother than at zero
strain: In zero strain, there is a strong first-order transition, while under strong $\langle
100 \rangle$ it becomes much broader. We hypothesize that this is a consequence of the uniaxial stress lifting
the degeneracy between the (100) and (010) directions.
\end{abstract}

\maketitle

\section{Introduction}

The magnetic order of \CAS{} offers a compelling example of how electronic order can cause a reduction in the point-group symmetry of the host lattice.  \CAS{} is a layered, tetragonal compound in
which a large-amplitude, incommensurate spin-density wave condenses at $T_N = 6.5$~K (in zero applied field)~\cite{Marcus18}. It is a heavy-fermion system, with a Kondo temperature of
$\sim$14~K~\cite{Seo12}. The magnetic order shows entropy balance with a Fermi liquid, showing that at $T_N$ the Ce spins are in fact incorporated into the Fermi sea through the Kondo
effect~\cite{Zhao16}.  For $c$-axis fields below $\sim$3~T, the in-plane wavevector of the spin-density wave (SDW) is, in reciprocal lattice units, either $(\eta, \eta, 1/2)$ or $(\eta, -\eta, 1/2)$, with $\eta \approx
0.136$~\cite{Marcus18}, and in selecting locally which of these two possibilities condenses, the point-group symmetry of the system is locally reduced from tetragonal to orthorhombic. 

However \CAS{} is highly polarizable under a $c$-axis magnetic field~\cite{Balicas05, Thamizhavel03}, and as field is applied the modulation amplitude of the SDW decreases. At $\sim$3~T there is a
first-order transition, above which the two components coexist microscopically. If their amplitudes are equal tetragonal symmetry is restored. The modulation persists up to $\sim$6~T, beyond which, at
another first-order transition, the system becomes uniformly polarized~\cite{Balicas05, Lorenzer13}. 

The association of weaker order with microscopic coexistence extends to other systems. The
correlated-electron material Sr$_3$Ru$_2$O$_7$ has a magnetically ordered phase with low-amplitude SDWs oriented along the $(100)$ and $(010)$ directions, which coexist microscopically~\cite{Lester15,
Brodsky17}. In at least some iron-based superconductors, single-component $C_2$-symmetric magnetic order at low dopings gives way, as the order is suppressed through doping towards its quantum
critical point, to two-component $C_4$-symmetric order~\cite{Avci14, Boehmer15}. In the rare-earth tritelluride compounds, which are weakly orthorhombic, charge density waves with perpendicular
wavevectors can coexist microscopically, however only when the amplitude of the dominant component is suppressed through chemical pressure~\cite{Ru08, Moore10}. An advantage of studying \CAS{} is that
the strength of the density wave order, and, apparently, the strength of competition between the two possible density wave components, can be tuned externally with magnetic field.

The transition at $\sim$3~T in \CAS{} is a single- to multi-component transition, and probably also an orthorhombic to tetragonal transition. The above-listed systems may have similar transitions,
driven by a tuning parameter such as doping or magnetic field. We hypothesize that the single- to multi-component transition in \CAS{} is first order because there is no natural pathway to tune
between the two phases, and further that an externally-applied symmetry-breaking field such as in-plane uniaxial stress could provide such a pathway and change the transition into a continuous
(i.e. second-order) transition. Under tetragonal lattice symmetry the two components are degenerate, and the two natural possibilities for $T$ well below $T_N$ are that strong competition
allows only one component to condense, yielding spontaneous symmetry breaking, or that weaker competition allows them to coexist microscopically with equal amplitude. A transition between these states
would be strongly discontinuous. However in-plane uniaxial stress would lift the degeneracy and may allow one component to dominate on both sides of a single- to multi-component transition, and the
amplitude of the other to grow continuously from zero.

Here, we test this hypothesis in two stages. First, we test whether the high-field phase (between $\sim$3 and $\sim$6~T) is in fact tetragonal. In principle, it is possible to have states with, in the
absence of any symmetry-breaking field, $|\Delta_{11}| \neq |\Delta_{1\bar{1}}|$ and both $|\Delta_{11}|$ and $|\Delta_{1\bar{1}}| \neq 0$ [where $\Delta_{11}$ and $\Delta_{1\bar{1}}$ are respectively
the amplitudes of the $(\eta, \eta, 1/2)$ and $(\eta, -\eta, 1/2)$ density waves]. However this would require a more delicate tuning of interactions--- more precisely, terms beyond fourth order in
a two-component Ginzburg-Landau theory.  We test for such symmetry breaking by ramping the applied uniaxial stress through zero (that is, between compressive and tensile). If there is spontaneous
symmetry breaking, there should be a first-order transition at zero stress, where the favored direction of the symmetry breaking flips. In our earlier study of \CAS{} under uniaxial stress and at zero
field~\cite{Park18}, such a transition was observed, corresponding to the transition between $\mathbf{q} = (\eta, \eta, 1/2)$ and $(\eta, -\eta, 1/2)$. However, here, to high
sensitivity no such transition is observed for the high-field phase, indicating that it is most likely tetragonal (i.e. $|\Delta_{11}| = |\Delta_{1\bar{1}}|$).

Second, we apply strong uniaxial compression and observe the evolution of the field-temperature phase diagram. We focus mostly on stress along $\langle 100 \rangle$ directions (that is, Ce-Ce bond
directions). Although the density wave components in unstressed \CAS{} are oriented along $\langle 110 \rangle$ directions and $\langle 100 \rangle$ stress is a transverse field to this order, stress
along $\langle 110 \rangle$ directions has a much smaller quantitative effect than $\langle 100 \rangle$ stress~\cite{Park18}.  Here, at the strongest applied $\langle 110 \rangle$ compression the
transition at $\sim$3~T broadens only slightly, and it is difficult to be confident that this observed change is intrinsic. In the previous study it was found that $\langle 100 \rangle$ compression by
$\sim$0.5\% almost certainly rotates the principal axes of the low-field magnetic order from $\langle 110 \rangle$ to $\langle 100 \rangle$. Therefore, we work primarily with $\langle 100 \rangle$
stress, and compressions of $\sim0.5$\% and higher.

\section{Summary of previous results}

\begin{figure}[t]
\includegraphics[width=8.6cm]{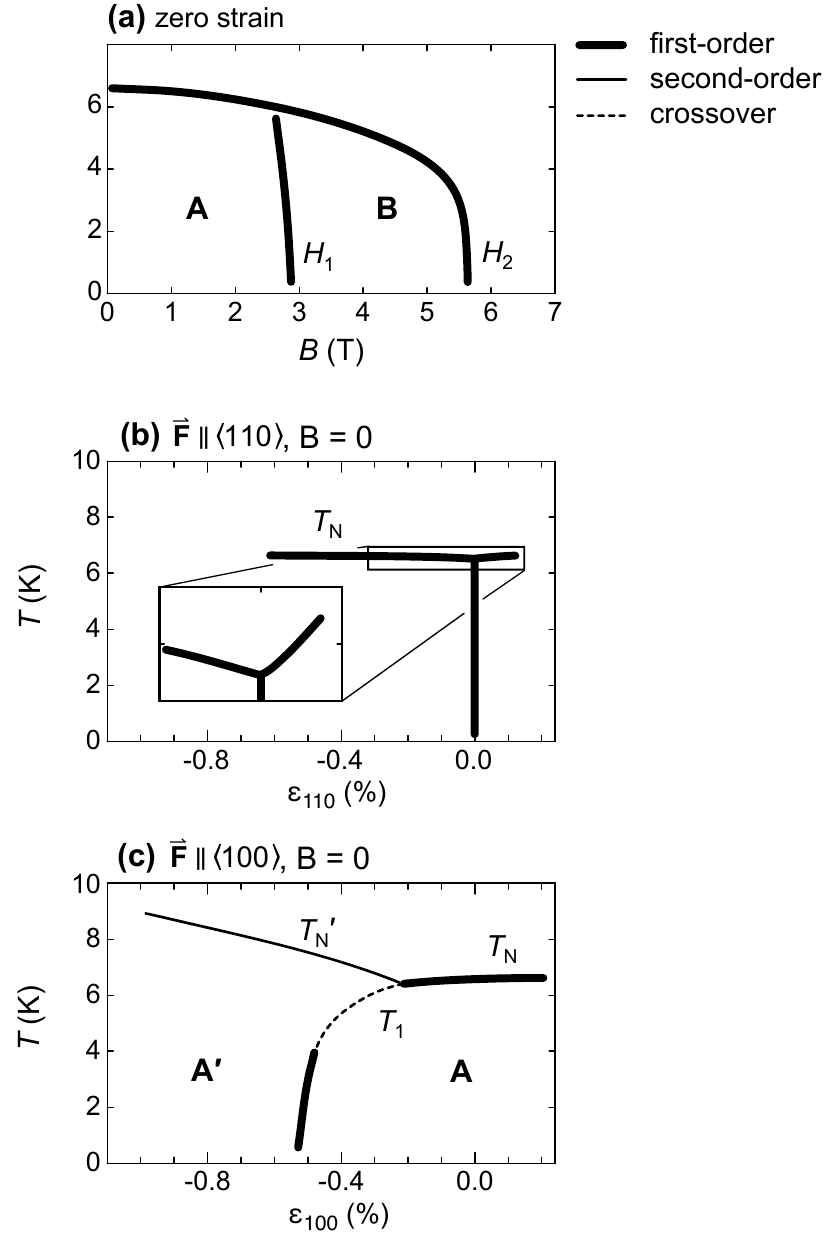}
\caption{\label{fig.1} (a) Field-temperature phase diagram of unstressed \CAS{}, for field along the $c$ axis. In Ref.~\cite{Zhao16} the near-vertical transition lines at $H_1$ and $H_2$ were found
to be first order, and in Ref.~\cite{Park18} the transition at $H=0$, $T=T_N$ was also found to be, probably, weakly first order. (b) and (c) Schematic $H=0$ strain-temperature phase diagrams, such as
they could be resolved in measurements, for pressure applied along a (b) $\langle 110 \rangle$ lattice direction, and (c) $\langle 100 \rangle$ direction.} 
\end{figure}

We begin by providing more detail, in Fig.~1, on previous results. The field-temperature phase diagram of
unstressed \CAS{} is shown in panel (a)~\cite{Zhao16, Balicas05}. The metamagnetic transitions occur at $\mu_0
H_1 = 2.8$~T and $\mu_0 H_2 = 5.9$~T. We label the low- and high-field phases the A and B phases. 

The magnetic order of \CAS{} has been shown to be sensitive to hydrostatic pressure, with a
modest pressure of $\sim$2~GPa inducing a new magnetic phase~\cite{Seo12}. It is similarly sensitive to
uniaxial stress applied along a $\langle 100 \rangle$, but not a $\langle 110 \rangle$, direction. In panel
(b) we show the stress-temperature phase diagram at zero field for stress applied along a $\langle 110
\rangle$ direction, inducing a longitudinal strain $\varepsilon_{110}$~\cite{Park18}.  (The sample is under
conditions of uniaxial stress, so there will also be transverse strains of the opposite sign, following the
sample's own Poisson's ratios) The line of first-order transitions along $\varepsilon_{110}=0$ shows that the
magnetic order spontaneously lifts the $(110)$/$(1\bar{1}0)$ symmetry of the $T > T_N$ lattice. However,
despite this notable qualitative effect, the quantitative effect of $\langle 110 \rangle$ stress on $T_N$ is
small.

In panel (c) we show the stress-temperature phase diagram for $\langle 100 \rangle$ stress. When \CAS{} is
compressed by more than $\sim$0.25\% along a $\langle 100 \rangle$ direction, the transition at $T_N$ splits
into two transitions, at temperatures $T_1$ and $T_N'$. The transition at $T_N'$ is second order. $T_1
\rightarrow 0$ at a compression of about 0.5\%, and we label the new high-strain phase A$'$. $\langle 100
\rangle$ strain is a transverse field with respect to the A phase, which has $\langle 110 \rangle$ principal
axes, so it is not surprising that $T_N$ varies only weakly with $\varepsilon_{100}$. The much stronger,
linear dependence of $T_N'$ on $\varepsilon_{100}$ is strong evidence that $\langle 100 \rangle$ strain is a
longitudinal field with respect to the A$'$ phase, in other words that the principal axes have rotated to the
$\langle 100 \rangle$ directions.  This would occur if, for example, the in-plane SDW wavevector rotated from
$(\eta, \pm \eta)$ to $(\eta', 0)$ or $(0, \eta')$. 

The transition between the A and A$'$ phases is first order below $\approx$4~K, and in resistivity data
appears to be a crossover above. However if different symmetries are broken in the A and A$'$ phases there
must be a true transition line between them.

\section{Methods}

CeAuSb$_2$ crystals were grown according to the methods described elsewhere~\cite{growth1, growth2}. They were oriented to a precision of $\sim$3$^{\circ}$ by Laue diffraction, then beams were cut
from the crystals with the long axis along a $\langle$100$\rangle$ or $\langle$110$\rangle$ direction. The crystals naturally grow in a plate-like geometry, however they were further polished in order
to obtain a uniform sample thickness. The samples were then mounted into a home-built uniaxial pressure apparatus~\cite{Hicks14} (using Stycast 2850FT epoxy), in which force is applied along their long
axis. The pressure apparatus is driven by piezoelectric actuators. It incorporates a displacement sensor placed in parallel with the sample. As in previous reports~\cite{Brodsky17, Park18}, we
estimate that $\sim$80$\%$ of the applied displacement is transferred to the central, exposed portion of the sample, with the rest going into deformation of the ends of the sample and the epoxy. In
other words, the strains reported here are the applied displacement divided by the exposed length of the sample, multiplied by 0.8, and we estimate a $\sim$20\% sample-to-sample error on this strain
determination.  

We measured a total of six samples, two cut along a $\langle$110$\rangle$ direction and four along a $\langle$100$\rangle$ direction. The first five samples were also studied in Ref.~\cite{Park18},
and have the same numbering here.

For the $\langle$110$\rangle$ samples, zero strain was taken as the location of the first-order transition: This transition is expected to occur at zero strain, there is no other feature that could
mark the neutral strain point, and, finally, taking it to mark zero implies a room temperature to 0~K thermal contraction for CeAuSb$_2$ of $\sim$~0.25$\%$, which is a typical value for a metal. For
the $\langle$100$\rangle$ samples, there is no feature in the response at the neutral strain point, so we took zero strain to be at the same applied displacement as for the $\langle 110 \rangle$
samples. 

The resistivity of \CAS{} changes strongly at the N\'{e}el transition and also across the metamagnetic transitions. In principle, when measuring resistivity under uniaxial stress the results should be
corrected for a geometric contribution, which is the change in resistance that would still be observed if the sample resistivity were held constant, due to the applied change in sample dimensions.
However, the resistivity of \CAS{} varies strongly with strain and we neglect this correction.

\begin{figure}[t]
\includegraphics[width=8.5cm]{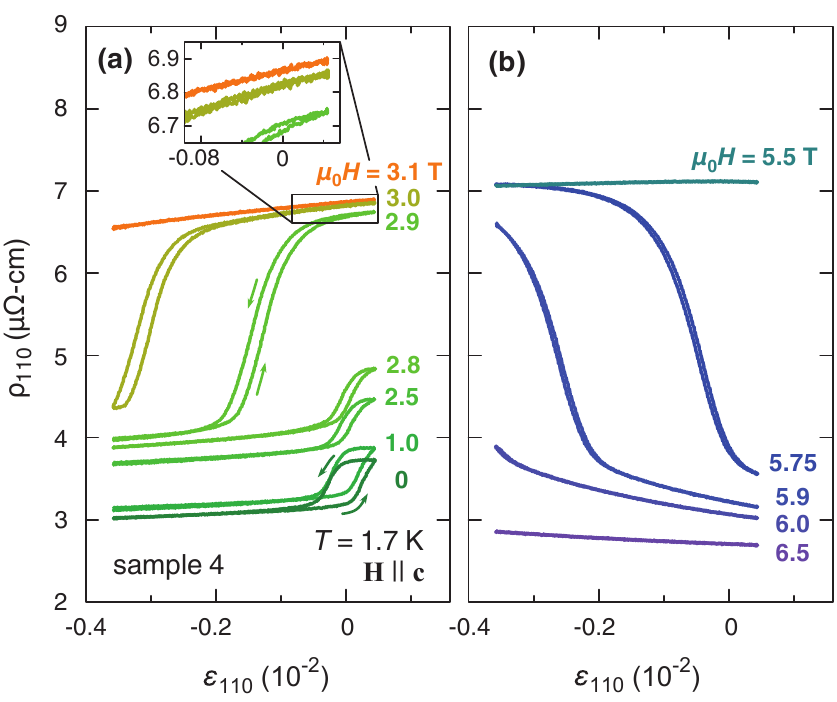}
\caption{(color online) \textbf{$\rho_{110}$($\varepsilon_{110}$) at selected magnetic fields.} $\rho_{110}$ is the resistivity measured along the length of the sample, cut along a $\langle 110
\rangle$ direction, and $\varepsilon_{110}$ in the longitudinal strain achieved through uniaxial stress applied along this sample. (a) The data taken between $H$~=~0 and 3.1~T. (b) The data taken between
$H$~=~5.5 and 6.5~T.} 
\end{figure}

\section{Results: testing the symmetry of the high-field phase}

Fig.~2 shows the results of ramping the stress applied along a $\langle 110 \rangle$ direction, at various fixed fields $H$ and $T = 1.7$~K.  Below $H_1$ (the transition field between the A and B
phases), the resistivity shows a step-like response across $\varepsilon_{110}$~=~0. There is clear hysteresis between the increasing- and decreasing-strain ramps. This is the first-order transition
between $(\eta, \eta, 1/2)$ and $(\eta, -\eta, 1/2)$ spin-density wave order, and the presence of this transition proves that the A phase is $C_2$-symmetric. The hysteresis shrinks as the field is
increased, implying a decreasing energy barrier for flipping domains. 

At fields around $H_1$ and $H_2$, there are first-order transitions at $\varepsilon_{110} \neq 0$, which correspond respectively to strain-driven transitions between the A and B phases, and between
the B phase and high-field paramagnetic phase. However, as shown in the inset of Fig.~2(a), to high precision there is no first-order transition across $\varepsilon_{110} = 0$ within the B phase. We
conclude that it most likely preserves symmetry between the $(110)$ and $(1\bar{1}0)$ directions.

\begin{figure}[t]
\includegraphics[width=85mm]{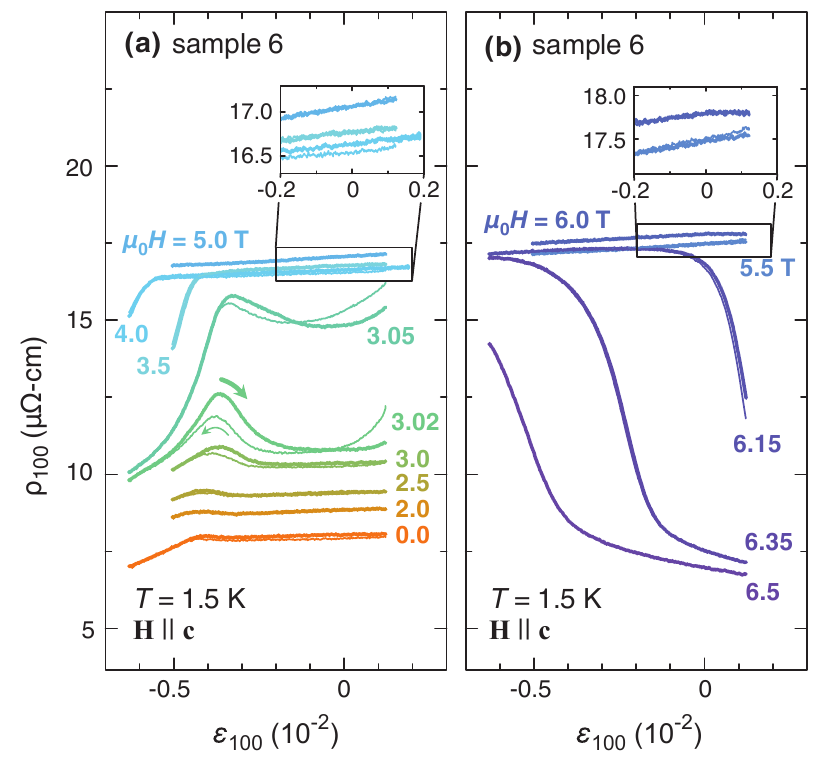}
\caption{(color online) \textbf{$\rho_{100}$($\varepsilon_{100}$) at selected magnetic fields.} $\rho_{100}$ is the resistivity measured along the length of the sample, cut along a $\langle 100
\rangle$ direction, and $\varepsilon_{100}$ in the longitudinal strain achieved through uniaxial stress applied along this sample. (a) The data taken between $H$~=~0 and $H$~=~5.0~T. (b) The data taken between
$H$~=~5.5 and 6.5~T.}
\end{figure}

Equivalent data for stress ramps along a $\langle 100 \rangle$ direction are shown in Fig.~3, with the temperature held constant at 1.5~K. For fields below 3~T, the transition between the A and A$'$
phases is visible as a sharp change in slope of $\rho(\varepsilon_{100})$, at $\varepsilon_{100} \sim -0.5$\%. The hysteresis that shows that this transition is first order is not visible in the
figure, however it was resolved in Ref.~\cite{Park18}. There also appears to be a similar transition for $\mu_0 H > 3$~T, where, again, $\rho$ is nearly strain independent for $|\varepsilon_{100}| < 0.5$\% but depends much more sensitively on $\varepsilon_{100}$ for $|\varepsilon_{100}| > ~0.5$\%. We will discuss this further below. Finally, for fields right in the vicinity of $H_1$,
the resistivity is strongly hysteretic.

\begin{figure}[t]
\includegraphics[width=85mm]{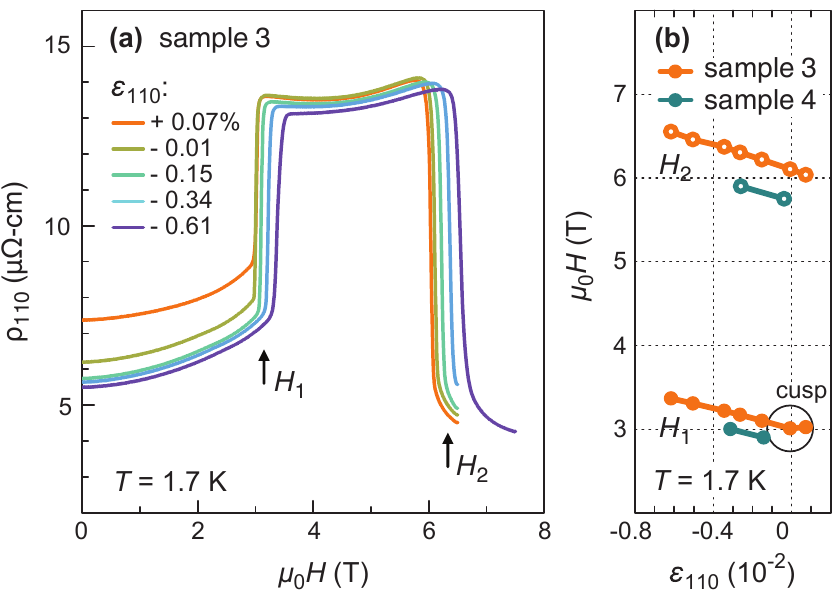}
\caption{\label{fig.2} (color online) (a) $\rho_{110}(H)$ at selected $\langle$110$\rangle$ strains, where the temperature was kept constant at $T$~=~1.7~K. (b) The $H-\varepsilon_{110}$ phase diagram
for $T$~=~1.7~K. Here, the $H_1$ and $H_2$ data points for sample 3 were taken from the field derivative maxima and the minima, whereas for sample 4, points were taken from the mid-point of the
step-like transition.} 
\end{figure}

\begin{figure*}[t]
\includegraphics[width=17cm]{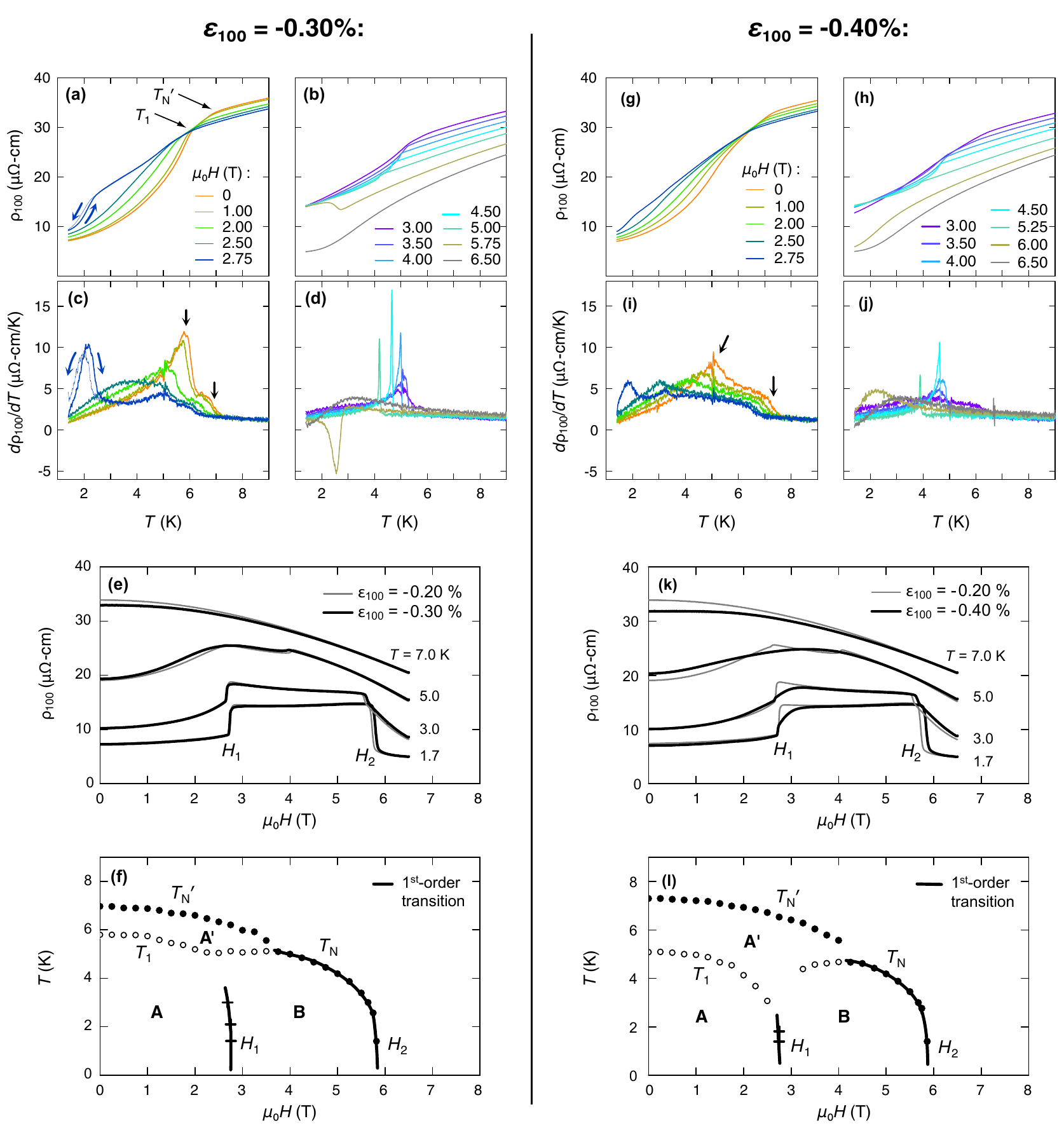}
\caption{\label{fig.5} (color online) Field-temperature phase diagrams under applied compression along a $\langle 100 \rangle$ direction. Data are from sample~2. \textbf{(a-f) 0.3\%
compression.} (a) and (b): $\rho_{100}(T)$ at selected magnetic fields for (a) $H<H_1$ and (b) $H>H_1$. (c) and (d): Derivatives $d\rho_{100}/dT$ of the curves in, respectively, panels (a) and (b).
(e) $\rho_{100}(H)$ at selected fixed temperatures, plotted together with, for reference, data at $\varepsilon_{100} = -0.2$~\%.  (f) The field-temperature phase diagram derived from the data in
panels (a-e). \textbf{(g-l) 0.4\% compression.} The panels mirror panels (a-f).}
\end{figure*}

For our present purpose, the important feature in Fig.~3 is that there is no apparent transition at $\varepsilon_{100} = 0$, either in the low-field or high-field phase. Our sensitivity to changes in
$\rho$ was $\delta \rho / \rho$~$\sim$~10$^{-4}$. We conclude that the magnetic order in both phases most likely preserves the symmetry between the $(100)$ and $(010)$ directions. Because the data in
Fig.~2 indicate that the B phase also preserves $(110)$/$(1\bar{1}0)$ symmetry, we conclude that it is probably $C_4$ symmetric, i.e. tetragonal.

In Ref.~\cite{Marcus18}, it was proposed that the B phase may possess a subtle symmetry-breaking between the $(100)$ and $(010)$ lattice directions. The neutron data are consistent both with a
multi-component checkerboard order, which preserves $(100)$/$(010)$ symmetry, and a ``woven'' order, which like the checkerboard order is a multi-component density wave, but in which the local pattern
of magnetization lifts $(100)$/$(010)$ symmetry. It was speculated that the woven order might be favored because it allows each component to have a larger amplitude while keeping the local maximum and
minimum magnetizations within a narrower range. In principle, because the symmetry breaking in the proposed woven order is subtle, its effect on resistivity could yet be below our resolution.  However
we conclude that it is more likely that the $B$ phase is in fact the tetragonal, checkerboard order.

\section{Results: large strains}

We start, in Fig.~4, with the results from large $\langle 110 \rangle$ strain, which essentially confirm the finding of Ref.~\cite{Park18} that the quantitative coupling of $\langle 110 \rangle$
orthorhombicity to the magnetic order is weak. The figure shows results of measurement of $\rho(H)$ at various fixed strains $\varepsilon_{110}$, and $T = 1.7$~K.  The form of $\rho_{110}(H)$, a
first-order increase at $H_1$ and first-order decrease at $H_2$, is familiar from previous studies~\cite{Zhao16, Balicas05, Lorenzer13}. Field sweeps reported in previous studies confirmed that these
are first-order transitions, with hysteresis. Both transitions move to modestly higher fields with $\langle 110 \rangle$ compression. The transition at $H_1$ broadens slightly as the sample is
compressed, and arguably, looking at the highest compression, slightly more than the transition at $H_2$. This broadening could be an early sign of the hypothesized stress-driven evolution from a
first-order to a continuous transition. However it could also be an extrinsic effect of stress gradients in the sample, due to minor bending of the sample as stress is applied.

As highlighted in Fig.~4(b), $H_1(\varepsilon_{110})$ shows a cusp at zero strain, consistent with the
above-described observations that the A phase lifts the (110)/(1$\bar{1}$0) symmetry of the lattice.  No such
cusp is apparent in $H_2$, consistent with our finding above that the high-field phase is probably
$C_4$-symmetric.

We move on to large $\langle 100 \rangle$ stress, for which the response is considerably richer. To understand
the evolution of the field-temperature phase diagram, we start in Fig.~5 with relatively low strains,
$\varepsilon_{100} = -0.3$ and $-0.4$\%, where there are observable changes in the phase diagram, but where
the connections with the zero-strain phase diagram also remain clear. Results for $\varepsilon_{100} = -0.3$\%
are shown in panels (a)-(f), and show that at this strain the field-temperature phase diagram is only
minimally altered from that at zero strain. As was reported in Ref.~\cite{Park18}, this is a large enough
strain to split the N\'{e}el transition (into the transitions at temperatures $T_1$ and $T_N'$). This splitting
is observable as two breaks in slope in the $\rho(T)$ curves shown in panel (a), and can also be seen as two
step-like features in the derivative $d\rho/dT$, shown in panel (c). The splitting persists, essentially
unchanged, for all fields $H<H_1$. 

At 2.75~T, right in the vicinity of $\mu_0H_1$, a very prominent first-order transition appears at $\sim$2~K.  This is the transition between the A and B phases, visible in a temperature ramp because
the transition line is not perfectly vertical in field-temperature space. 

Separate transitions at $T_1$ and $T_N'$ remain visible at $H_1$. As $H$ is further increased [see panels (b) and (d), which show the data for $H>H_1$], the splitting decreases and the transitions
merge at $\mu_0 H \approx 3.6$~T. At higher fields the transition into the B phase occurs, as at zero strain, through a single, first-order transition. The first-order nature of the transition is
apparent in the very sharp peaks in $d\rho_{100}/dT$, in panel (d).

To identify any changes in $H_1(T)$ and $H_2(T)$, field ramps were performed at constant temperature, with the
results shown in panel (e). They are shown together with data from field ramps at $\varepsilon_{100} =
-0.2$\%, which match the zero-strain data very well~\cite{Zhao16}. The increase in strain from -0.2 to -0.3\%
induces very little change; the only substantial qualitative change is that at $T = 5$~K there is no longer an
identifiable transition at $H_1$.

\begin{figure}[t]
\includegraphics[width=85mm]{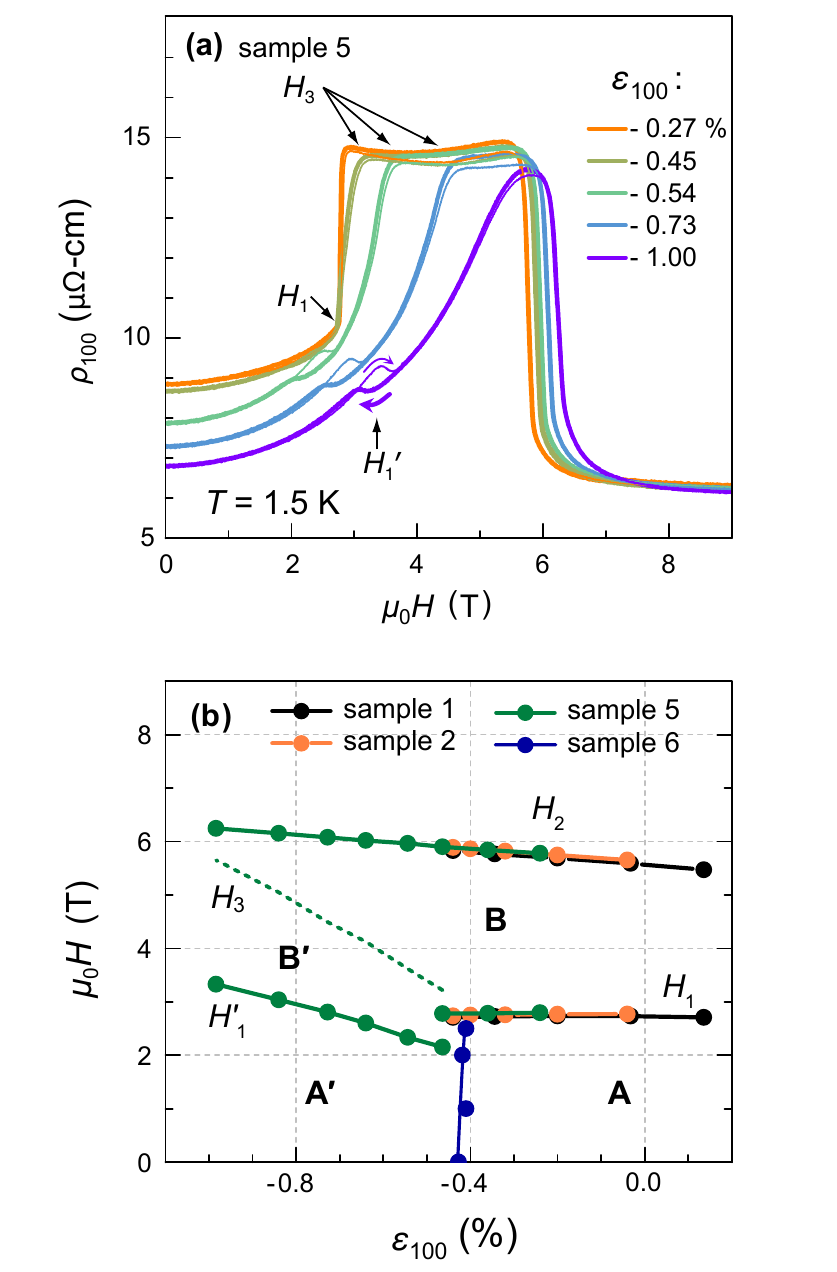}
\caption{(a) $\rho_{100}(H)$ at selected fixed strains $\varepsilon_{100}$ and $T=1.5$~K. (b)$H-\varepsilon_{100}$ phase diagram at $T = 1.5$~K.}
\end{figure}

Putting all this data together, we obtain a field-temperature phase diagram for $\varepsilon_{100} = -0.3$\%,
shown in panel (f). It is qualitatively similar to the field-temperature phase diagram of the unstressed
sample, except in a narrow band along the $T_N(H)$ line where the A$'$ phase has appeared.

In panels (g)-(l) we show the equivalent data for a higher strain, $\varepsilon_{100} = -0.4$\%. The data are
qualitatively similar to those at $\varepsilon_{100}=-0.3$\%, though they show stronger effects from the applied
lattice orthorhombicity. Most obviously, the A$'$ phase now occupies a larger region of the phase diagram.
Another prominent change is that the first-order transition between the A and B phases has become considerably
weaker. In panel (e), where $\varepsilon_{100} = -0.3$\%, the first-order step in $\rho$ constitutes almost
the entire transition, while in panel (k), where $\varepsilon_{100} = -0.4$\%, the first-order step is only a
small feature in a transition that overall has become broad and rounded.

Fig.~6 presents data at stronger compressions. $\rho(H)$ is shown for various fixed strains $\varepsilon_{100}$ in panel (a). As the A phase is fully suppressed and replaced even at $T \rightarrow 0$
by the A$'$ phase, the transition at $H_1$ disappears completely. A new first-order transition appears, at a field that we label $H_1'$. The similarity of $H_1$ and $H_1'$ indicate that the physical
process driving these transitions is likely to be similar; we hypothesize that both are single- to multi-component transitions. However the form of the transition at $H_1'$ is different from that at
$H_1$: the hysteresis is much wider, and the change in $\rho$ across the transition is much smaller. It is also apparent in Fig.~6(a) that whereas $H_1$ is nearly independent of $\varepsilon_{100}$,
$H_1'$ varies rapidly with $\varepsilon_{100}$. The strain dependence of $H_1$ and $H_1'$ can also be seen in the strain-temperature phase diagram for $T = 1.5$~K, shown in panel (b).  It is a similar
situation to $T_N$ and $T_N'$, where the former is nearly invariant with $\varepsilon_{100}$ while the latter has a strong linear dependence, and constitutes further evidence that the
principal axes have rotated to the $\langle 100 \rangle$ directions in the A$'$ phase.

Two further features apparent in Fig.~6(a) should be noted. One is that the resistivity varies much more strongly with $\varepsilon_{100}$ in the A$'$ than the A phase: $\rho(H)$ for $H<H_1$ changes
very little with $\varepsilon_{100}$ for $|\varepsilon_{100}|<0.5$\%, but varies much more rapidly at larger compressions. Another is that a set of transition fields that we label
$H_3$ has appeared in the high-field phase. Below $H_3$, $\rho$ varies strongly with both field and $\varepsilon_{100}$, while above it is nearly strain- and field-independent. In the phase diagram
of Fig.~6(b), we identify $H_3$ as a transition line into a high-strain, high-field phase that we label B$'$. This transition is also visible in the $\rho(\varepsilon_{100})$ data shown in Fig.~3(a). As a summary of our data, we present in Fig.~7 a three-dimensional field-strain-temperature phase diagram.

\begin{figure}[t]
\includegraphics[width=85mm]{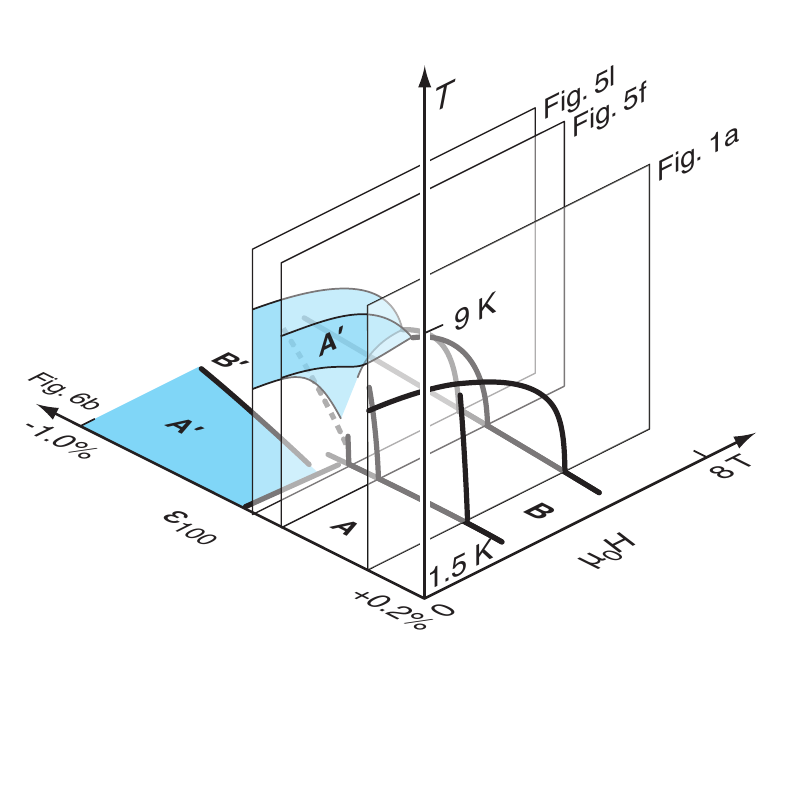}
\caption{A summary of our results: the field-strain-temperature phase diagram of CeAuSb$_2$ under applied $\langle 100 \rangle$ uniaxial pressure. As discussed in the text, Phases A and B are single- and multi-component SDW orders, in which the components propagate along $\langle 110 \rangle$ directions. Phases A$'$ and B$'$ are probably also single- and multi-component SDW orders, however in which the components propagate along $\langle 100 \rangle$ directions.}
\end{figure}

\section{Discussion}

We have presented evidence that the B phase, in contrast to the A phase, does not lift the $C_4$ symmetry of the lattice of \CAS{}. We have also shown that the field-temperature phase diagram of
\CAS{} has a rich dependence on applied $\langle 100 \rangle$ orthorhombicity. At a minimum, we have identified the strains and fields where transitions in the magnetic order occur. Definitive
identification of how the magnetic order changes under strain will require further measurements, for example neutron scattering on uniaxially stressed \CAS{}. We proceed in our discussion with
inferences that might be drawn based on the resistivity alone.

The $H$-$\varepsilon_{100}$ phase diagram of Fig.~6(b) is essentially a 2$\times$2 grid, where the small-$|\varepsilon_{100}|$ phases are A and B, and the large-$|\varepsilon_{100}|$ phases are A$'$ and
B$'$. Neutron scattering data~\cite{Marcus18} have shown that the A and B phases are respectively single- and multi-component orders. From the quantitative similarity of $H_1$ and $H_1'$, the
transition field between the A$'$ and B$'$ phases, we hypothesize that the A$'$ and B$'$ phases are also, respectively, single- and multi-component orders.

Both the A$'$ and B$'$ phases are marked by a much stronger sensitivity to applied $\langle 100 \rangle$ lattice orthorhombicity than the A and B phases. The resistivity varies much more strongly with
$\varepsilon_{100}$ in both the A$'$ and B$'$ phases than in the A and B phases, and the transition fields that bound the A$'$ and B$'$ phases, $H_1'$ and $H_3$, vary more rapidly with $\varepsilon_{100}$
than those bounding the A and B phases, $H_1$ and $H_2$. We therefore conclude that $\langle 100 \rangle$ orthorhombicity is a longitudinal field for both the A$'$ and B$'$ phases, where it is a
transverse field for the A and B phases; in other words the density wave components comprising both the A$'$ and B$'$ phases have $\langle 100 \rangle$ principal axes. The 2$\times$2 grid therefore
appears to comprise the possible combinations of single- and multi-component order, and $\langle 110 \rangle$ and $\langle 100 \rangle$ principal axes.

It is interesting that the B phase is, apparently, so insensitive to $\varepsilon_{100}$. In the neutron
study~\cite{Marcus18}, strong scattering peaks were observed at wavevectors $(2\eta, 0, 0)$ and $(0, 2\eta,
0)$, which were interpreted as results of nonlinear mixing of the $(\eta, \pm \eta, 1/2)$ components. If we
interpret the $\varepsilon_{100}$-independence of $\rho$ within the B phase as indicating that the magnetic
order is similarly unaffected by $\varepsilon_{100}$, then our observations support this interpretation: the
magnetic order, until the boundary with the B$'$ phase is reached, is essentially independent of
$\varepsilon_{100}$ because $\langle 100 \rangle$ orthorhombicity is a transverse field to the two fundamental
components, and the peaks at $(2\eta, 0, 0)$ and $(0, 2\eta, 0)$ are interference peaks, not independent
components that couple directly to $\langle 100 \rangle$ orthorhombicity. 

We conclude by returning to our original hypothesis that applied lattice orthorhombicity, by selecting a
preferred direction, would change the first-order transition at $H_1$ into a continuous transition. The
high-strain data partially but not completely support this hypothesis. Instead of a dominant first-order step
at $H_1$, under large $|\varepsilon_{100}|$ the field evolution of the resistivity, and by inference the
magnetic order, is overall more gradual. However there is still a first-order transition, at $H_1'$. The
change in $\rho$ across $H_1'$ is small, suggesting that the change in magnetic order is minimal. It is
possible that higher-order interactions still drive the single- to multi-component transition to be first-order, and
the weaker component onsets with small but non-infinitesimal amplitude.

\section*{Acknowledgment}
The authors thank O. Erten for the fruitful discussions and Paul Canfield and Veronika Fritsch for their assistance with sample growth. We acknowledge the financial support of the Max Planck Society.
JP acknowledges the financial support of the National Research Fondation of Korea (NRF) funded by the Ministry of Science and ICT (Grant No. 2016K1A4A4A01922028). HS acknowledges the financial
support of PRESTO, JST (Grant No. JPMJPR16R2) and Grant-in-Aid for Young Scientists (Grant No. 16H06015). Raw data for all figures in this paper are available at \url{http://edmond.mpdl.mpg.de/imeji/collection/_n8DQUqmqC4qrw7x} or \url{https://dx.doi.org/10.17617/3.1o}.

\end{document}